\documentclass[aps,prb,reprint]{revtex4-1}

\usepackage{amsmath,amssymb} % mathematics
\usepackage{amsfonts} % mathematics
\usepackage{bm} % bold in math-mode
\usepackage{here}
\usepackage{url}
\usepackage[pdftex]{graphicx} % figure
\usepackage{braket} % braket
\usepackage{physics}
\begin{document}

%%%%%%%%%%%%%%%%%%%%%%%%%%%%%%%%%%%%%%%%%%%%%%%%%%%%%%%%%%%%
%%%%%%%%%%%%%%% title and author information %%%%%%%%%%%%%%%
%%%%%%%%%%%%%%%%%%%%%%%%%%%%%%%%%%%%%%%%%%%%%%%%%%%%%%%%%%%%
% Title
\title{Exact matrix product states at the quantum Lifshitz tricritical point 
in a spin-1/2 zigzag-chain antiferromagnet with anisotropic $\Gamma$-term}

% Author information
\author{Hidehiro Saito}
\email{saito-hidehiro722@g.ecc.u-tokyo.ac.jp}
\author{Chisa Hotta}
\email{chisa@phys.c.u-tokyo.ac.jp}
\affiliation{Department of Basic Science, University of Tokyo, Meguro-ku, Tokyo 153-8902, Japan}
\date{\today}

%%%%%%%%%%%%%%%%%%%%%%%%%%%%%%%%%%%%%%%%
%%%%%%%%%%%%%%% abstract %%%%%%%%%%%%%%%
%%%%%%%%%%%%%%%%%%%%%%%%%%%%%%%%%%%%%%%%
\begin{abstract}
Quantum anisotropic exchange interactions in magnets can induce 
competitions between phases in a different manner from those typically driven by geometrically frustrated interactions. 
We study a one-dimensional spin-1/2 zigzag chain with such an interaction, $\Gamma$-term, 
in conjunction with the Heisenberg interactions. 
We find a ground state phase diagram featuring a multicritical point 
where five phases converge: 
a uniform ferromagnet, two antiferromagnets, Tomonaga-Luttinger liquid and 
a dimer-singlet coexisting with nematic order. 
This multicritical point is simultaneously quantum tricritical and Lifshitz, and 
most remarkably, it hosts multi-degenerate ground state wave functions, whose exact form 
is obtained in the matrix product form and its degeneracy increases in squares of system size. 
\end{abstract}

% Title
\maketitle

%%%%%%%%%%%%%%%%%%%%%%%%%%%%%%%%%%%%%%%%%%%%
%%%%%%%%%%%%%%% Introduction %%%%%%%%%%%%%%%
%%%%%%%%%%%%%%%%%%%%%%%%%%%%%%%%%%%%%%%%%%%%
{\it Introduction.} 
Traditionally, phase diagrams for quantum magnetism in one dimension (1D) are quite simple, 
predominantly characterized by the Tomonaga-Luttinger liquid (TLL) 
which sometimes undergoes a transition to a magnetic phases breaking discrete symmetries like Ising N\'eel order. 
The introduction of geometrical frustrations can lead to 
emergent quantum disordered phases 
%with lattice-symmetry breaking, 
such as dimer-singlet phase on a zigzag chain\cite{Majumdar1969,Majumdar1969-2,Haldane1982,Okamoto1992,Eggert1996,White1996} 
and the nematic phase on frustrated ladders\cite{Hikihara2008,Sudan2009}. 
These states are sufficiently trivial in the context of quantum topology, 
as they are approximated by product-state wave functions. 
Still, a realization of Majumdar-Gosch(MG) state\cite{Majumdar1969,Majumdar1969-2} 
inside the dimer-singlet phase may deserve particular emphasis, 
because it stands as one of a handful of exact ground states in realistic quantum many-body models. 
\par
Recently, quantum anisotropic exchange interactions are added as another ingredient to enrich quantum magnetism. 
An iconic example is the Kitaev interaction that serves as a source of spin liquid 
with long-range entanglement and topological excitations\cite{Kitaev2006,Nasu2015,Kasahara2018}. 
In reality, the Kitaev interaction cannot avoid coexistence with the $\Gamma$ and Heisenberg terms 
that largely restrict the phase space of spin liquid\cite{Jackeli2009,Chaloupka2010}. 
However, a very rich phase diagram with spiral, stripe or other spatially modulated phases 
\cite{Rau2014-kitaev} 
show that they provide strong frustration or competition in a way 
not easily attained by the geometrical frustration effect. 
\par
The anisotropic exchange interactions can naturally arise in the Mott insulating state of 4d, 5d, and 4f electrons 
with strong spin-orbit couplings like iridium oxides, iridates, and rare-earth magnets\cite{WitczakKrempa2014,Rau2016}. 
When derived microscopically, they are classified into 
three categories; Kitaev-type bond-oriented diagonal exchange, 
$\Gamma$-term representing bond-symmetric, off-diagonal (different spin component) exchange, 
and the bond-antisymmetric Dzyaloshinskii-Moriya exchange interactions. 
For rare-earth-based materials with high crystal symmetric octahedral ligands, 
the SU(2) symmetric Heisenberg interactions are dominant and a small $\Gamma$-term adds on that top\cite{Rau2014,Saito2023}. 
In this Letter, we clarify the role of $\Gamma$-term by simplifying it to an idealized form 
in a 1D zigzag Heisenberg chain with geometrical frustration. 
Although $\Gamma$-term was previously a less important secondary term in the Kitaev magnetism, 
our ground state phase diagram turns out to be rich, as it includes the multicritical point 
showing both Lifshitz and tricritical nature
\cite{Lifshitz1942,Hornreich1975,Nelson1975,Abdel-Hady1996,Chauhan2019}. 
\par
The Lifshitz point is a finely adjusted point in a phase diagram 
at which a uniform phase and a spatially modulated ordered phase meet the disordered phase 
\cite{Michelson1977,Hornreich1975}. 
In the early days, it was discussed in the ANNNI model\cite{Selke1988}. 
Similarly, the multicritical point in metamagnets is explained by the competing antiferromagnetic  
and spin flopped states in a magnetic field\cite{Nelson1974,Fisher1974,Nelson1975,Kosterlitz1976} 
or in liquid $\,^4$He\cite{Liu1973}. 
These examples are governed by two competing Ising order parameters 
that originate from the explicit magnetic anisotropies of the Hamiltonian. 
By contrast, in our case, the compatibility of the $\Gamma$-term and the discrete symmetry of the crystal lattice 
spontaneously selects the magnetic hard axis, generating three Ising order parameters. 
The three are decoupled in the Landau theory 
and will form a Gaussian fixed point similar to the bicritical/tetracritical point of the metamagnets\cite{Nelson1974,Fisher1974}. 
Our tricritical Lifshitz point stands out because it occurs 
at zero temperature and meet the quantum disordered phase, 
it fall exactly on the MG line with exact singlet states, 
and most intriguingly, its multi-degenerate ground states turn out to be exactly 
described by a matrix product state (MPS) representation. 

%*%*%*%*%*%*%*%*%*%*%*%*%*
%      fig1
%*%*%*%*%*%*%*%*%*%*%*%*%*
\begin{figure}
    \centering
    \includegraphics[width=8.5cm]{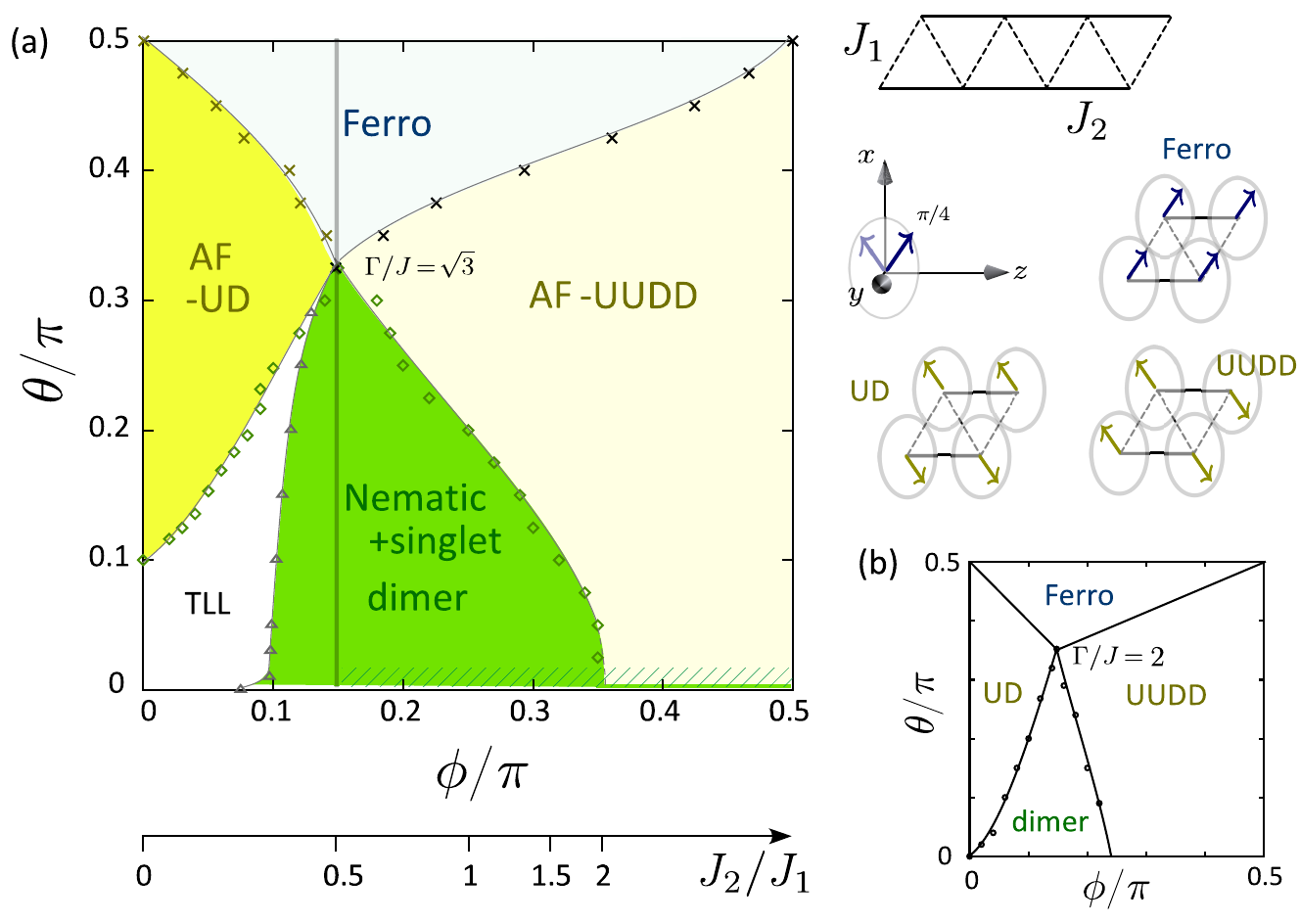}
    \caption{
    (a) Ground state phase diagram of Eq.(\ref{eq:ham}) obtained by DMRG. 
     Solid line $J_2/J_1=1/2={\rm arctan}(\phi)$ is the Majumdar-Gosch line 
      where the singlet product state is the exact eigenstate. 
     The multicritical point is located on that line at $\Gamma/J=\sqrt{3}$. 
%% $(\Gamma/J, J_2/J_1)=(\sqrt{3},1/2)$. 
     The right panels show the zigzag chain and the global $xyz$ coordinate, 
     where UD and UUDD have $(1,1,0)$ and the Ferro phase has $(1,-1,0)$ as a spontaneously emergent hard axis of magnetic moment, respectively. 
    (b) Mean-field phase diagram where the multicritical point appears at $(\Gamma/J, J_2/J_1)=(2,1/2)$. 
    The data points are the mean-field solution from the bond operator approach based on the dimer. 
    }
 \label{f1}
\end{figure}
%*%*%*%*%*%*%*%*%*%*%*%*%*
%*%*%*%*%*%*%*%*%*%*%*%*%*
%      fig2
%*%*%*%*%*%*%*%*%*%*%*%*%*
\begin{figure}[t]
    \centering
    \includegraphics[width=8cm]{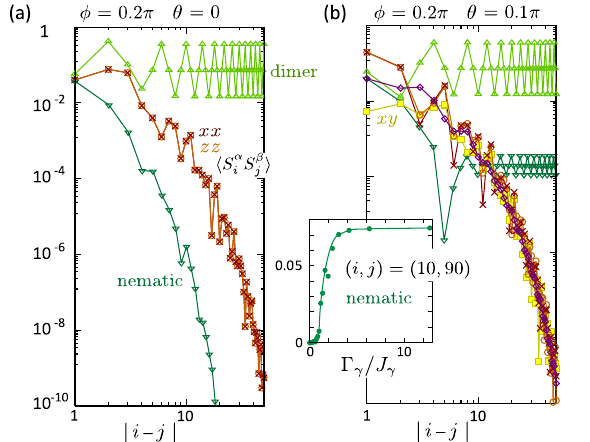}
    \caption{(a) Two-point correlation functions $\langle O_iO_j\rangle$ 
of spins $O_i=S_i^{\alpha_i}$ of $(\alpha_i\alpha_j)=xx,zz$, and $xy$, 
dimer $O_i=\bm S_i\cdot\bm S_{i+1}$ and nematic operators $O_i=S^x_i S^y_{i+1} + S^y_i S^x_{i+1}$ 
obtained by DMRG with $N=100$ and OBC. 
    (a) $\theta=0$ and (b) $\theta=0.1\pi$ with $\phi=0.2\pi$. 
    Inset shows the evolution of nematic correlation at $(i,j)=(10,90)$ with $\Gamma_\gamma/J_\gamma$. 
}
    \label{f2}
\end{figure}
%*%*%*%*%*%*%*%*%*%*%*%*%*
%*%*%*%*%*%*%*%*%*%*%*%*%*
%      fig3
%*%*%*%*%*%*%*%*%*%*%*%*%*
\begin{figure*}
    \centering
    \includegraphics[width=18cm]{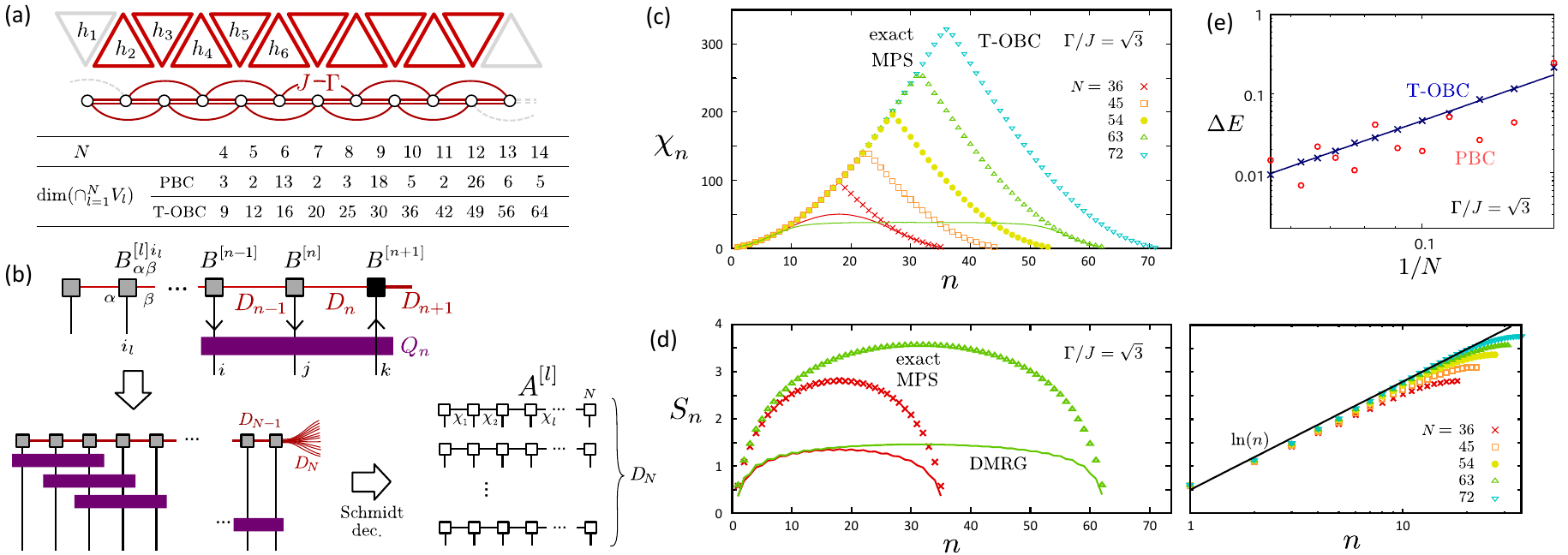}
    \caption{(a) Illustration of the model at $J_1=2J_2=J, \Gamma_1=2\Gamma_2=\Gamma$ 
    as a sum of $h_l$ with $l=2,\cdots,N-1$ (T-OBC), in comparison with the 1D description, 
    and the degeneracy of the ground state $D_{N}$ up to $N=14$ for PBC and T-OBC. 
    (b) Protocol to construct an exact MPS with T-OBC. 
    Matrix $B^{[n+1]}$ is successively generated from the $n$- and $(n-1)$-th matrix elements 
    by solving linear equations represented by $Q_n$. 
    The matrix elements of $B^{[l]}$ ($l=1,\cdots,N$) are separated into 
    a set of $D_{N}$ independent degenerate MPS $\{A^{[l]}\}$. 
   (c) Bond dimensions $\chi_n$ of exact MPS $A^{[l]}$ 
    for $N=36,45,54,63,72$ 
    which conforms to three-site translation. 
   (d) Entanglement entropy $S_n$ of the exact MPS with $N=36,63$ and T-OBC 
   about the bipartition to $n$ and $(N-n)$-site subsystems as a function of $n$. 
   Data are averaged over the degenerate ground states. 
   Right panel gives the  entanglement scaling: Solid line is $S_n \sim (c/3) \ln n$ with $c=3$. 
   Solid lines in panels (c) and (d) are the corresponding DMRG results 
   with numerically accurate energy $Ne_0$. 
   (e) Energy gap $\Delta E$ between the ground state $Ne_0$ and the first excited states 
       obtained by ED for T-OBC and PBC. Solid line shows finite size scaling behavior $\Delta E\propto N^{-2.60}$. 
}
    \label{f3}
\end{figure*}
%*%*%*%*%*%*%*%*%*%*%*%*%*
\par
{\it Model and phase diagram.} 
We consider a quantum spin-1/2 Hamiltonian on a zigzag chain given as 
\begin{equation}
{\cal H}= \sum_{j} \sum_{\gamma=1,2} 
J_\gamma \bm S_j \cdot \bm S_{j+\gamma} 
+ \Gamma_\gamma (S^x_jS^y_{j+\gamma}+ S^y_{j}S^x_{j+\gamma}), 
 \label{eq:ham}
\end{equation}
where $J_\gamma$ and $\Gamma_\gamma$ are the Heisenberg and anisotropic exchange interactions between 
nearest ($\gamma=1$) and next nearest ($\gamma=2$) spins 
for which we take site indices alternatively between legs. 
For later convenience, we parameterize them as 
$\Gamma_1/J_1=\Gamma_2/J_2=\tan\theta$ and $J_2/J_1=\Gamma_2/\Gamma_1=\tan\phi$, 
where the range $0\le \phi \le \pi/2$ corresponds to antiferromagnetic $J_1$ and $J_2$\cite{Okunishi2003,Okunishi2008,Hikihara2010}. 
The sign of $\Gamma_\gamma$ does not matter as it is erased by the local unitary transformation. 
We take the spin quantization axis as the one shown in Fig.~\ref{f1}; 
the $z$-axis is taken parallel to the chain and 
the $xy$-plane has an O(2) symmetry where we set 
$y$-axis perpendicular to the triangular plane for convenience. 
\par
We show in Fig.~\ref{f1}(a) the ground state phase diagram of Eq.(\ref{eq:ham}) obtained 
by the density matrix renormalization group (DMRG) calculation\cite{White1992}. 
Here, we take typically $N=100$ lattice sites and adopt two cases, 
open boundary condition (OBC) and a sine-square deformation (SSD), 
where the latter can accurately evaluate incommensurate ordering and suppresses finite-size effects
\cite{Nishino2011,Hotta2012,Hotta2013}. 
The $\Gamma_{\gamma}=0$ ($\theta=0$) limit is the zigzag $J_1$-$J_2$ model 
where the transition from a TLL phase to the dimer singlet phase occurs at $J_2/J_1\sim 0.241$
\cite{Majumdar1969-2,Haldane1982,Okamoto1992,Eggert1996,White1996}. 
At $\Gamma_{\gamma}\ne 0$, the two phases extend and undergo a second order transition 
to the antiferromagnetically ordered (AF) phases 
which we call UD and UUDD for $J_2/J_1\le 0.5$ and $\ge 0.5$, respectively.  
The AF phases experience first-order transitions to the Ferromagnetic (F) phase. 
The magnetic moments of the two AF phases are locked parallel to $(x,y,z)=(1,1,0)$ 
and that of F phase to $(1,-1,0)$. 
Similar spin orientation is observed in the 1D Heisenberg chain with a bond alternative $\Gamma$-terms\cite{Yang2020}. 
\par
The phase diagram has a few remarkable features. 
First, the three second-order transition lines separating 
UD, TLL, dimer singlet, and UUDD meet at the single point, $(J_2/J_1, \Gamma/J)=(1/2, \sqrt{3})$. 
It is the endpoint of the two first-order transition lines separating the F and UD, UUDD phases. 
This point is tricritical because the three order parameters are adjusted as we see shortly. 
\par
Second, we find that $J_2/J_1=\Gamma_2/\Gamma_1=1/2$ is an MG line; the singlet product state, 
$|\Psi_{\rm MG} \rangle=\prod_{j=1}^{N/2} |s_j\rangle$ with 
$|s_j\rangle=(|\!\uparrow \downarrow\rangle-|\!\downarrow\uparrow\rangle)/\sqrt{2}$ on $[2j-1,2j]$ sites, 
continues to be the exact eigenstate of Eq.(\ref{eq:ham}). 
\par
Finally, at $\Gamma\ne 0$ and off the MG line, 
the nematic order emerges inside the dimer-singlet phase at all finite values of $\Gamma$. 
Figure~\ref{f2} shows two-point correlation functions of magnetic and nonmagnetic operators obtained by DMRG. 
At $\Gamma=0$, all the magnetic correlations decay in power law with distances 
and the singlet-singlet correlation function stays dominant. 
When $\Gamma >0$, a nematic-nematic correlation starts to sustain at $|i-j|\rightarrow \infty$, 
whose value continuously increases from a zero value at $\Gamma=0$. 
\par
{\it Multicriticality.} 
We first clarify the underlying competition of phases near the multicritical point. 
To illustrate it, we prepare the basis set of site $j$ 
using the up and down spin states about the global $xyz$-coordinate, 
$|\!\!\uparrow\rangle,|\!\!\downarrow\rangle$, as
$|u\rangle_j=|\!\!\uparrow\rangle_j + e^{i\pi/4}|\!\!\downarrow\rangle_j$, 
$|d\rangle_j=|\!\!\uparrow\rangle_j - e^{i\pi/4} |\!\!\downarrow\rangle_j$, 
$|f\rangle_j=|\!\!\uparrow\rangle_j + e^{i3\pi/4}|\!\!\downarrow\rangle_j$, 
and its conjugate $|f^*\rangle_j$. 
Their magnetic moment points in the $(\pm 1,\mp 1,0)$ and $(\pm1,\pm1,0)$ directions, respectively. 
The variational scheme considering a maximally four-site-period naturally provides 
a product state $\prod_{j=1}^N |\psi_j\rangle$ where $|\psi_j\rangle$ is expanded as 
a linear combination of the above-mentioned four sites. 
The lowest energy state gives mean-field solutions which are, 
$|\Psi_{\rm UUDD}\rangle= \prod_{j=1}^{N/4} |u\rangle_{4j-3} |u\rangle_{4j-2} |d\rangle_{4j-1} |d\rangle_{4j}$, 
$|\Psi_{\rm UD}\rangle= \prod_{j=1}^{N/2}|u\rangle_{2j-1} |d\rangle_{2j}$, and 
$|\Psi_{\rm F} \rangle= \prod_{j=1}^{N} |f\rangle_{j}$, 
whose energies are 
$E_{\rm UUDD}=-(J_2+\Gamma_2)N/4$, 
$E_{\rm UD}=(-J_1-\Gamma_1+J_2+\Gamma_2)N/4$, and 
$E_{\rm F}=(J_1-\Gamma_1+J_2-\Gamma_2)N/4$, respectively. 
These three states are exclusive to each other and meet at $\Gamma/J=2$ 
in mean-field phase diagram in Fig.~\ref{f1}(b); 
it is slightly off the exact multicritical point, 
while successfully capturing the essense of the numerical phase diagram. 
Here, we have added the data points derived from the bond-operator mean field approach\cite{Sachdev1990,Ueda2007} 
based on dimers, which will be discussed elsewhere. 
\par
The implications of the mean-field results become distinct 
when we rotate the spin coordinate by $\pi/4$ about the $z$-axis; 
The Hamiltonian in the new coordinate $x'y'z'$ is given as 
${\cal H}^{\pi/4} = {\cal H}_{x'y'} + {\cal H}_{z'}$, 
and by dropping off ${\cal H}_{z'}$, 
we obtain ${\cal H}^{\pi/4}\sim {\cal H}_{x'y'}$ with 
\begin{eqnarray}
{\cal H}_{x'y'} =\sum_{j=1}^N \sum_{\gamma=1,2} 
(J_\gamma-\Gamma_\gamma) S^{x'}_jS^{x'}_{j+\gamma} 	
,+(J_\gamma + \Gamma_\gamma) S^{y'}_jS^{y'}_{j+\gamma}, 
\label{eq:ham45} 
\end{eqnarray}
where ${\cal H}_{z'} = \sum_{j=1}^N\sum_{\gamma=1,2} J_\gamma S^{z'}_{j}S^{z'}_{j+\gamma}$ 
turned out to be irrelevant in the renormalization analysis because of the quantum fluctuation. 
The U and D in the mean-field solutions are the 
Ising type solution of Eq.(\ref{eq:ham45}) in the $x'$-directions, 
and F is the one in the $y'$-direction, whose energies cross at the multicritical point. 
\par
As another way of understanding this spontaneous selection of magnetic hard axes, 
one may go back to Eq.(\ref{eq:ham}) finding that it is invariant under the 
$\pi$-rotation about both the $(1,1,0)$ and $(-1,1,0)$ axes, 
which transforms the spins as $(S^x,S^y,S^z)\rightarrow (S^y,S^x,-S^z)$ and $(-S^y,-S^x,-S^z)$, respectively. 
These two directions are thus robust against the quantum fluctuation 
and hence are encoded in Eq.(\ref{eq:ham}) as invisible hard axes. 
\par
The two or more competing Ising states remind of a bicritical/tetracritical point 
of ANNNI model\cite{Selke1988} and metamagnets\cite{Nelson1974,Fisher1974,Nelson1975,Kosterlitz1976}. 
Their origin, the competing ferromagnetic and antiferromagnetic Ising exchanges in the former, 
and the magnetic hard axis and the transverse magnetic field in the latter  
are visible in their Hamiltonian. 
Their multicritical phase transition at finite temperature occurs due to the thermal entropic effect. 
In contrast, our Ising anisotropy comes from the interplay of quantum 
anisotropic exchange and the lattice symmetry, and 
the quantum disorder due to quantum fluctuation drives the multicriticality at zero temperature. 
\par
%*%*%*%*
{\it Exact solutions.} When $J_1=2J_2\equiv J$ and $\Gamma_1=2\Gamma_2=\Gamma$, 
the couplings of diagonal rungs are doubled from those of legs, 
and the Hamiltonian Eq.(\ref{eq:ham}) can be rewritten as the sum of local Hamiltonians, 
$h_l$, of the $l$-th triangular unit based on $[l+1,l,l-1]$-th spins 
as ${\mathcal H} = \sum_{l} h_l$ with 
\begin{eqnarray}
h_l= \sum_{i,j \in l} J \bm{S}_i \cdot \bm{S}_j
 + \sum_{i,j \in l} \Gamma \left(S_i^x S_j^y + S_i^y S_j^x \right) , 
\label{eq:tr}
\end{eqnarray}
which is shown schematically in Fig.~\ref{f3}(a). 
At $\Gamma=0$, the lowest energy of $h_l= J (\bm{S}_{l-1} + \bm{S}_l + \bm{S}_{l+1})^2/2 - 9J/8$ 
is $\epsilon_0=-3J/4$, attained by a total $S=1/2$ for all triangle which is the MG state. 
When $\Gamma > 0$, $J$ and $\Gamma$ terms in Eq.(\ref{eq:tr}) commute 
and the MG singlet state remains as the exact eigenstate of Eq.(\ref{eq:ham}). 
Notice that the $J_{\gamma}$ and $\Gamma_{\gamma}$ terms of Eq.(\ref{eq:ham}) {\it do not commute}. 
\par
At the multicritical point, $\Gamma/J=\sqrt{3}$, 
the ground state is highly degenerate including a doubly degenerate MG state. 
This is already unusual because conventional critical ground states typically lack such degeneracy. 
Even more exceptional is that the degrees of degeneracy and {\it a full set of exact solutions} are accessible. 
To prove this, we present two methodologies: 
the first involves solving a set of linear equations satisfying conditions to have the energy $E=-N\epsilon_0$, 
while the second entails iteratively constructing a matrix product state. 
In these methods, we employ a triangular open boundary condition (T-OBC) shown in Fig.~\ref{f3}(a); 
Suppose that we have a set of degenerate ground states of ${\mathcal H} = \sum_{l=1}^N h_l$ 
under a periodic boundary condition (PBC). 
Even if we eliminate the two triangles, $h_1$ and $h_N$, as illustrated in Fig.~\ref{f3}(a), 
all the degenerate ground states persist as those of ${\cal H}=\sum_{l=2}^{N-1} h_l$ 
with energy $-(N-2)\epsilon_0$. 
\par
{\it Solving linear equations.}\;\; 
There are eight eigenstates of a triangular unit in Eq.(\ref{eq:tr}), which are the four ``nematic" states with $S=3/2$, 
$|\psi_{n\Uparrow}^\pm\rangle$, $|\psi_{n\Downarrow}^\pm\rangle$, 
and four magnetic states, $|\psi_{m\Uparrow}\rangle$, $|\psi_{m\Uparrow}^*\rangle$, 
$|\psi_{m\Downarrow}\rangle$, $|\psi_{m\Downarrow}^*\rangle$ with $S=1/2$. 
Using 0/1 representing up/down spins on $[l+1,l,l-1]$ sites in the descending order, they read
\begin{align}
&|\psi_{n\Uparrow}^\pm \rangle =\sqrt{3} |000\rangle \pm i(|011\rangle+|101\rangle+|110\rangle), \\ 
&|\psi_{m\Downarrow}\rangle= |011\rangle+\omega|101\rangle+\omega^2|110\rangle ,
\label{eq:trst}
\end{align}
with $\omega=e^{i2\pi/3}$, 
where $\Uparrow$ states are obtained from the $\Downarrow$ ones
by converting the spins as $0\leftrightarrow 1$. 
Among them, six are the constituents of the ground state, 
$\{|\psi_l^{\rm gs}\rangle\} \equiv 
\{|\psi_{n\Uparrow}^-\rangle, |\psi_{n\Downarrow}^+\rangle, |\psi_{m\Uparrow}\rangle, 
|\psi_{m\Downarrow}^*\rangle, |\psi_{m\Uparrow}\rangle, |\psi_{m\Downarrow}^*\rangle\}$. 
Because triangles share edges, 
the entire ground state needs to entangle this six basis in a nontrivial manner. 
Most straightforwardly, we can project out the two excited states on each triangle. 
Let the subspace of the Hilbert space be given as 
$V_l=\{ \{|\psi_l^{\rm gs}\rangle\}\otimes |m\rangle_{j=1,\cdots,2^{N-3}}\}$, 
where $\{|j\rangle\}$ is the subspace spanned by the $N-3$ sites that do not belong to the $l$-th triangle. 
All the states on $\cap_{l=1}^N V_l$ are the ground state $|\Psi\rangle$; 
they satisfy two conditions per triangle as $ Q |\Psi\rangle =0 $ with 
\begin{align}
& Q =\oplus_{l=1}^N Q_l, \label{eq:linear}
\\
%% 000=0 001=1 010=2 011=3 
%% 100=4 101=5 110=6 111=7  
& Q_l=\left(\begin{array}{rrrrrrrr}
\sqrt{3} & 0 & 0& -i & 0 & -i & -i & 0 \\
0 &i & i&  0 & i & 0 & 0 & \sqrt{3}
\end{array}\right), 
\label{eq:amat}
\end{align}
where the $(i=1,\cdots, 2^3)$-th rows of $Q_l$ operates on the states of the $l$-th triangle 
in the bit representation of $i$, namely 
$|000\rangle,\;|001\rangle,\;\cdots,|111\rangle$, we applied in Eq.(\ref{eq:trst}). 
The number of linear equations is $N\times 2^{N-2}$. 
We numerically confirmed that solving them gives a full set of 
ground state by comparing with the full exact diagonalization (ED) of the original Hamiltonian, 
while the practically available size is limited to $N\lesssim 20$. 
However, the number of degeneracy of the ground states is 
$D_{N}={\rm dim} (\cap_{l=1}^N V_l) = 2^N-{\rm rank} Q$, 
which is obtained iteratively by 
${\rm dim}(\cup_{i=1}^{n} V_i)/{\rm dim}(\cup_{i=1}^{n-1} V_i)=\frac{n+4}{2(n+2)}$ (odd $n$)
and $\frac{n+5}{2(n+3)}$ (even $n$) and we find, 
\begin{align}
D_{N}=\left\{ 
\begin{array}{ll}
(N+2)^2/4 & ({\rm even}\;\;N)  \\ 
(N+1)(N+3)/4 \;\;&  ({\rm odd}\;\;N). 
\end{array}
\right.
\label{eq:bond}
\end{align}
The actual values of $D_{N}$ are presented in Fig.~\ref{f3}(a). 
\par
{\it Exact MPS solutions.}\;\;
The derivation of an MPS\cite{Fannes1992} is practically important for accurately 
representing quantum states with large $N$\cite{Schollwck2011}. 
Usually, the MPS of the critical state is only approximately available, 
whereas ours provides a full set and the exact description. 
The protocol to obtain the MPS is shown schematically in Fig.~\ref{f3}(b). 
Suppose that we have an $n$ set of matrices that starts from the left edge matrix as 
$B^{[l]}=B_{\alpha\beta}^{[l]i_l}$ with $l=1,\cdots, n$; 
the $l$-th matrix has dimension $D_{l-1}\times D_{l}\times 2$ 
with $i_l=0/1$ spin degrees of freedom. 
For small $n$, we can obtain $\{B^{[l]}\}$ by a Schmidt decomposition of the exact wave function of ED. 
In increasing sites from $n$ to $n+1$, 
we consider a vector $\bm v$ whose $m=i+2j+4k$ element is $(B^{[n-1]i}B^{[n]j}B^{[n+1]k})$, 
that includes $D_n$ unknowns of $\alpha_{n+1}$-th column of $B^{[n+1]k}$. 
We have $D_{n-2}$ such vector for each choice of the row of $B^{[n-1]i}$. 
Therefore, $\bm v$ needs to fulfill $D_{n-2}$ different pairs of linear equations 
$Q_n \bm v = 0$ using Eq.(\ref{eq:amat}). 
Solving $2D_{n-2}$ simultaneous equations, we are able to construct at most $2D_{n}$ different solutions for 
$D_{n}$ unknowns, which altogether form $B^{[n+1]k}$. 
At $n=N$, we are left with $D_{N}$ free bonds on the r.h.s. 
By separating them into connected elements, 
we obtain $D_{N+1}$ different MPSs; 
We left-normalize each MPS to make them orthogonal with each other, 
and perform a Schmidt decomposition to reduce the bond dimension to discard the zero-weight ones. 
The final form $\{A^{[n]}\}_{n=1}^N$ of $\chi_{n}\times \chi_{n+1}\times 2$, 
serves as a full set of ground states. 
These operations are done by implementing the routines of iTensor\cite{Fishman2022}. 
\par
Figure~\ref{f3}(c) shows the bond dimension $\chi_N$ of $\{A^{[n]}\}_{n=1}^N$ 
for choices of $N$ that conform to the triangle-based translational symmetry. 
The $n\le N/2$ ones follow the exact values for all choices of $N$ and are symmetric about the center. 
The entanglement entropy (EE) $S_n$ about the bipartition to $n$ and $N-n$ 
obtained by averaging over $D_{N}$ ground states is shown in Fig.~\ref{f3}(d). 
When plotted using $\ln n$ they apparently extaporates to the critical behavior 
of $S_n\sim c\log(n)/3$\cite{Vidal2003,Calabrese2004} with central charge $c=3$ 
similarly to spin bose-metal\cite{Sheng2009}. 
This value is interpreted as the number of free bosonic excitations that represent 
the three adjacent magnetic phases that meet at the critical point. 
We parallelly performed the DMRG calculation for T-OBC and compared it with exact MPS. 
Because DMRG can only capture one of the degenerate ground states 
and favor minimally entangled states, the product MG state likely dominates over others, 
resulting in a plateau of $\chi_n$ and much smaller EE. 
Our exact MPS does not lose any essential information and has an advantage over it. 
\par
We finally confirm in Fig.~\ref{f3}(e) that the excitation energy 
$\Delta E$ above the ground state obtained by ED 
vanishes at $N\rightarrow\infty$ in power of $1/N$, confirming the gapless nature of the 
multicritical point. 
\par
{\it Conclusions.}\;\;
We discovered a Lifshitz tricritical point in a ground state phase diagram of the 
spin 1/2 zigzag ladder, and its exact MPS representation in a system of size $N$. 
Over fifty years of its history, such multicritical points have been revisited several times. 
Recent ones are possibly realized in materials. 
In NbFe$_2$, there exists a competition between ferromagnet and spin-density waves, 
and a magnetic field drives the transition down to low temperature 
toward the tricritical point\cite{Friedemann2018}. 
In Cu$_2$OSeO$_3$ that hosts skyrmions, 
several magnetic phases compete and exhibit both the tricritical point and 
the Lifshitz point separately in the magnetic phase diagram\cite{Chauhan2019}. 
Our study is motivated by a 4f magnet of spin-1/2 zigzag chain, YbCuS$_2$\cite{Ohmagari2020,Hori2022,Hori2023,Saito2023}; 
The neutron diffraction measurement reports an incommensurate magnetic structure at $T<1$k 
and an NMR suggests a gapless nonmagnetic excitation. 
However, the value of $\Gamma/J <0.1$ remains too small to reach a multicritical point in a laboratory. 
\par
From the theory point of view, there had been only a few examples of exact MPS states, 
AKLT\cite{Affleck1987,Schollwck2011,Fannes1992}, W-state\cite{Dr2000,Klimov2023} and Greengerger-Horne-Zeilinger state\cite{Greenberger2007,Caves2002}, which are simple and mostly related to idealized models or quantum computation. 
The present multicritical MPS adds rich and useful playground 
in a physically meaningful model in the condensed matter field. 

%%%%%%%%%%%%%%%%%%%%%%%%%%%%%%%%%%%%%%%%%%%%%%%%
%%%%%%%%%%%%%%% Acknowledgements %%%%%%%%%%%%%%%
%%%%%%%%%%%%%%%%%%%%%%%%%%%%%%%%%%%%%%%%%%%%%%%%
{\it Acknowledgments.}
We thank Tomotoshi Nishino, Frank Pollmann, Shunsuke Furuya for discussion, 
and Takahiro Onimaru, Chikako Moriyoshi, Kenji Ishida, Shunsaku Kitagawa, 
and Fumiya Hori for communications. 
This work is supported by the "The Natural Laws of Extreme Universe" (No. JP21H05191) KAKENHI for Transformative Areas from JSPS of Japan, and JSPS KAKENHI Grants No. JP21K03440.

%%%%%%%%%%%%%%%%%%%%%%%%%%%%%%%%%%%%%%%%
%%%%%%%%%%%%%%% Appendix %%%%%%%%%%%%%%%
%%%%%%%%%%%%%%%%%%%%%%%%%%%%%%%%%%%%%%%%
\appendix

%%%%%%%%%%%%%%%%%%%%%%%%%%%%%%%%%%%%%%%%%%%%
%%%%%%%%%%%%%%% bibliography %%%%%%%%%%%%%%%
%%%%%%%%%%%%%%%%%%%%%%%%%%%%%%%%%%%%%%%%%%%%
\bibliography{zigzag_ladder_ref}
\bibliographystyle{apsrev4-1}

\end{document}